\documentclass[10pt,preprint2]{aastex}

\def\HI{\mbox{\ion{H}{1}}}
\def\HII{\mbox{\ion{H}{2}}}

\def\MH{{{\rm H}_2}}

\def\dim#1{\mbox{\,#1}}
\def\figdir{.}

\begin{document}


\title{Redshifted 21 cm Emission From the Pre-Reionization Era\\
 I. Mean Signal and Linear Fluctuations}

\author{Nickolay Y.\ Gnedin}
\affil{CASA, University of Colorado, Boulder, CO 80309, USA;
gnedin@casa.colorado.edu}
\and
\author{Peter A.\ Shaver}
\affil{ESO, Karl-Schwarzschild-Strasse 2, Garching, D-85748, Germany;
pshaver@eso.org}

\begin{abstract}
We use cosmological simulations of reionization to predict the possible
signal from the redshifted $21\dim{cm}$ line of neutral hydrogen in the
pre-reionization era and to investigate the observability of this signal.
We show that the separation of the mean (global) signal over the whole
sky from the known foreground contamination may be feasible, but very
challenging. In agreement with previous studies, we demonstrate that
measuring angular fluctuations in the $\HI$ signal is likely to be
extremely difficult if not impossible because of the overwhelming
contamination from the galactic and extragalactic foregrounds. 
However,
we show that the sharp $\HI$ fluctuations in the frequency domain should be
easily separable from the relatively smooth spectra of the foregrounds,
and that these fluctuations should be detectable even at
moderate angular resolution ($\sim10-20$ arcmin).
\end{abstract}

\keywords{cosmology: theory - cosmology: large-scale structure of universe
  - diffuse radiation - galaxies: formation - galaxies: intergalactic
  medium - radio lines: general}

\section{Introduction}

The importance of the $21\dim{cm}$ line of neutral hydrogen has long been
recognized for its potential in studying the reionization of 
the universe. This possibility has been
emphasized many times in a large volume of previous research, beginning
with the pioneering work of Sunyaev \& Zel'dovich (1972), through the first
application of this idea to the modern scenario of galaxy formation by
Scott \& Rees (1990), to the recent detailed investigations (Madau,
Meiksin, \& Rees 1997; Gnedin \& Ostriker 1997; Shaver et al.\ 1999; Tozzi
et al. 2000; Di Matteo et al.\ 2002; Iliev et al.\ 2002; Ciardi \& Madau
2003; Iliev et al.\ 2003; Furlanetto, Sokasian, \& Hernquist 2004; Oh \&
Mack 2004, Zaldarriaga, Furlanetto, \& Hernquist 2004). 

The shape and strength of the $21\dim{cm}$ signal is ultimately connected
to the process of reionization of the universe. As numerical simulations of
reionization (Gnedin 2000; Ciardi, Stoehr, White 2003; Sokasian et al.\
2003; Ciardi, Ferrara, \& White 2003) become more detailed and precise, it
becomes possible to construct detailed reionization histories of the
universe that are consistent with recent observational data such as the
Lyman-alpha flux decrement in the IGM up to $z \sim 6$, and which, therefore,
can be used to make quantitative predictions about the expected
$21\dim{cm}$ signal. 

The difficulty of measuring the $21\dim{cm}$ signal is not so much its
weakness as the overwhelming contamination from galactic and
extragalactic foregrounds (Shaver et al.\ 1999; Di Matteo et al.\ 2002; Oh
\& Mack 2004) at these frequencies ($\nu \sim 100 - 200\dim{MHz}$). So the
question of the observability of the signal - which includes both the mean
signal over the whole sky and fluctuations in both the angular and
frequency domains - is the question of separating the cosmological signal
from the foreground contamination.

In this paper we consider the widespread fluctuations in the $21\dim{cm}$
signal caused by the (linear) density fluctuations in the cosmic
gas. Additional fluctuations in the $21\dim{cm}$ signal can also appear in
special regions, such as individual $\HII$ regions around possible rare
bright high redshift quasars (Madau et al.\ 1997); this topic will be the
subject of a forthcoming paper. But even if the signal from the vicinities
of high redshift quasars is more easily measurable, observations of quasars
do not substitute for the search for linear density fluctuations. Observing
signatures of quasars in the redshifted $21\dim{cm}$ signal will tell us
important information about the very first AGNs in the universe, but they
will not provide the kind of information that can be extracted from the
measurement of linear fluctuations. 

For example, measuring the amplitude of the linear density fluctuations at
$z\sim10$ provides information about the evolution of the matter power
spectrum at redshifts intermediate between those probed by the CMB
($z\sim1000$) and those probed by galaxy surveys ($z\sim1$). Such
information will be important in placing complimentary constraints on
cosmological parameters and the evolution of the cosmic equation of state.

\section{Simulations}

Since our goal is to investigate the observability of the redshifted
$21\dim{cm}$ line from the early universe, we need a plausible model for
the evolution of the physical state of the cosmic gas at that epoch. We use
numerical simulations with the Softened Lagrangian Hydrodynamics (SLH) code similar to those reported in Gnedin
(2000). The advantage of using the SLH code is that it incorporates all of
the important physical processes that affect the reionization of the
universe and the emission and absorption in the $21\dim{cm}$ line:
\begin{description} 
\item[Dark matter] is followed using the adaptive
  Particle-Particle-Particle-Mesh (P$^3$M) algorithm (Gnedin \&
  Bertschinger 1996).
\item[Gas dynamics] is followed on a quasi-Lagrangian deformable mesh
using the SLH algorithm.
\item[Star formation] is included using the Schmidt law in resolution
elements that sink below the numerical resolution of the code.
\item[Ionization and thermal balance] of the hydrogen and helium plasma
are followed exactly using a two-level implicit scheme.
\item[Molecular hydrogen] formation and destruction are followed
exactly (including radiative transfer effects) in the limit when the
fraction of hydrogen in the molecular form is small and the self-shielding
of $\MH$ is unimportant. (The latter is always the case in the simulation
presented in this paper because the numerical resolution is not sufficient
to resolve the formation of molecular clouds.) 
\item[Radiative transfer] is treated self-consistently in a 3D
  spatially-inhomogeneous and time-dependent manner using the Optically
  Thin Variable Eddington Tensor (OTVET) approximation of Gnedin \& Abel
  (2001). 
\end{description}

The SLH code also incorporates two important physical effects that were
emphasized by Madau, Meiksin, \& Rees (1997): heating by X-ray halos around
galaxies and heating by secondary electrons. The first effect is included
automatically by virtue of using the gas dynamical simulation, while the 
second effect is incorporated explicitly in the code as explained in
(Ricotti, Gnedin \& Shull 2003). 

Our knowledge of the evolution of the early universe is sketchy at
best. Therefore, in order to cover a sufficient parameter space, both in
terms of the cosmological parameters and possible variations in the
relative importance of different physical processes, we have run several
different sets of cosmological simulations. 

Star formation is incorporated in the simulations using a phenomenological
Schmidt law, which introduces two free parameters: the star formation
efficiency $\epsilon_{\rm SF}$ (as defined by eq.\ [1] of Gnedin 2000) and
the ionizing radiation efficiency $\epsilon_{\rm UV}$ (defined as the
energy in ionizing photons per unit of the rest energy of stellar
particles). 

Each set of simulations includes several runs with different values of
these parameters, and two pieces of observational data are used to
constrain the parameters. The star formation efficiency $\epsilon_{\rm SF}$
is chosen so as to normalize the global star formation rate in the
simulation at $z=4$ to the observed value from Steidel et al.\ (2001), and
the ultraviolet radiation efficiency $\epsilon_{\rm UV}$ is constrained by
the condition that the simulation reproduces the mean transmitted flux in
the Lyman-alpha forest at $z\approx6$ as measured by White et al.\ (2003).

\begin{table}
\caption{Simulation Parameters\label{sim}}
\begin{tabular}{lccccc}
\tableline
Set &
$\Omega_{m}$ &
$h$ &
$n$ &
Box size & Resolution \\
\tableline
\tableline
A & 0.27 & 0.71 & 1.0  & 4$h^{-1}\dim{Mpc}$ & 1$h^{-1}\dim{kpc}$ \\
B & 0.35 & 0.70 & 0.95 & 4$h^{-1}\dim{Mpc}$ & 1$h^{-1}\dim{kpc}$ \\
C & 0.35 & 0.70 & 0.97 & 4$h^{-1}\dim{Mpc}$ & 1$h^{-1}\dim{kpc}$ \\
D & 0.35 & 0.70 & 0.95 & 2$h^{-1}\dim{Mpc}$ & 1$h^{-1}\dim{kpc}$ \\
\tableline
\end{tabular}
\end{table}
Parameters of the four sets of simulations that we use in this paper are
given in Table \ref{sim}. All simulations include $128^3$ dark matter
particles, an equal number of baryonic cells on a quasi-Lagrangian moving
mesh, and about 3 million stellar particles that form continuously during
the simulation\footnote{Stellar particles sample the phase space density of
  the stellar component; they do {\it not\/} represent individual stars or
  star clusters.}, with the exception of the set D, in which only $64^3$
dark matter particles and baryonic cells are used - the latter set includes
several simulations with varied physical modeling, as explained below,
which makes it impractical to include a larger number of resolution
elements. The nominal spatial resolution of simulations is fixed at
$1h^{-1}$ comoving kpc, with the real resolution being a factor of two
lower. 

In all cases a flat cosmology is assumed, with $\Omega_{\Lambda,0} =
1-\Omega_{m,0}$, and normalization of the primordial fluctuations is
determined either from the {\it WMAP} data (Spergel et al.\ 2003) for set
A, or from the {\it COBE} data (White \& Bunn 1995) for the other three
sets. 

Given a simulation, we calculate the relative populations of
hyperfine levels of neutral hydrogen, taking into account the effects that
determine the level populations at high redshift: Lyman-alpha pumping and
collisions with both electrons and hydrogen atoms (Tozzi et al.\ 2000). We
then generate synthetic spectra at different points on the simulated sky,
and average them appropriately (depending on the observation modeled),
which provides us with the most accurate computation of the emission
signal, including velocity focusing.

\begin{figure}[t]
\plotone{\figdir/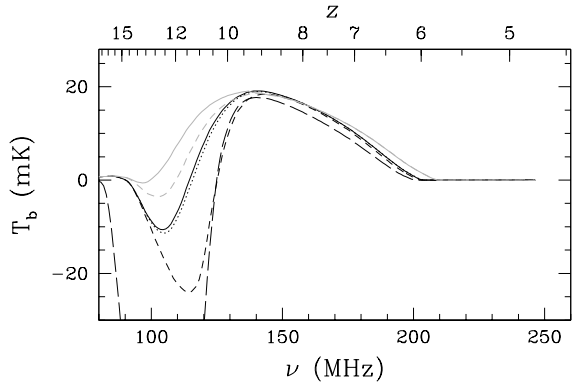}
\caption{\label{figME}The mean (over the whole sky) excess brightness
  temperature as a function of frequency and redshift for several
  simulations from set D: the fiducial model (black solid line), the same
  model with no heating by Lyman-alpha recoil (black dotted line), with no
  Lyman-alpha heating and no radiation above 4Ry (black short-dashed line),
  and with no Lyman-alpha heating, no radiation above 4Ry, and no shock
  heating (black long-dashed line). Two gray lines show the fiducial model
  with sources of ionization having a power-law quasar-like spectrum rather
  than a stellar-type spectrum. The solid gray line shows the model without
  secondary electrons included, while the gray dashed line shows the same
  model with secondary electrons included. All models are normalized to fit
  the observed evolution of the mean transmitted flux in hydrogen
  Lyamn-alpha line, as explained above.
}
\end{figure}

It is interesting to look at the physical effects that might control the
characteristics of the redshifted $21\dim{cm}$ emission. Simulations are
the ideal tool to make such a study, because they offer complete control
over their physical ingredients, as is demonstrated in Figure \ref{figME}. 

Our fiducial model includes all relevant physical effects, and assumes that
sources of ionization have a stellar spectrum (see Gnedin 2000 for
details). The prediction of the fiducial simulation from set D that best
fits the mean transmitted Lyman-alpha flux data at $z\sim6$ for
the mean (averaged over the whole sky) $21\dim{cm}$ signal is shown in
Figure \ref{figME} with a black solid line. 

There are several features in this curve. The curve starts to deviate from
zero at $z\sim15$ - this is the moment when the first sources of
Lyman-alpha radiation begin to form in the simulation. The Lyman-alpha
photons couple the level population of the hyperfine transition in neutral
hydrogen to the kinetic temperature of the gas (c.f.\ Tozzi et al. 2000)
and drive them out of equilibrium with the Cosmic Microwave Background
(CMB). Because initially most of the gas in the universe is colder than the
CMB, the signal is seen as absorption against the CMB. By $z\sim12$, the
mean gas temperature has however risen above the CMB temperature (the
specific source of this heating is discussed in the following paragraphs),
so the $21\dim{cm}$ signal switches from absorption to emission. At
$z\sim10$ the ionization fronts in our fiducial model start to expand in
the low density IGM, which occupies most of the volume of the universe and
at $z\sim10$ still contains most of the mass. This process results in a
progressively larger fraction of neutral hydrogen being ionized and a
progressively lower $21\dim{cm}$ signal, until, finally, the universe is
fully reionized by $z\sim6$ (in accord with the SDSS observations, White et
al.\ 2003), and the $21\dim{cm}$ signal becomes unmeasurably small. 

The characteristic transition from absorption to emission at
$z\sim11-12$ is controlled by the heating of the cosmic gas. What is the
primary source of gas heating? Using the simulations from the set D, we
have investigated the role of various physical effects on the shape of the
cosmological $21\dim{cm}$ signal. The advantage of using simulations is
that we can switch various physical effects on and off at will. For
example, the dotted line in Fig.\ \ref{figME} shows the brightness
temperature evolution for the simulation, in which we do not include the
heating by recoil of Lyman-alpha photons in the temperature
equation for the cosmic gas. As one can see, the effect of Lyman-alpha
recoil is not significant in our fiducial model. 
This conclusion fully agrees with Chen \& Miralda-Escud\'e
(2003). 

In the further exploration of importance of various physical effects, we
show with the short-dashed line in Fig.\ \ref{figME} out fiducial model,
but with both Lyman-alpha heating and heating by X-rays excluded. i.e.\ all
sources are assumed to emit no radiation above $4\dim{Ry}$. 
As can be seen, X-rays have
a substantial effect on the evolution of the excess brightness temperature
at $15>z>9$ (in this particular cosmological model). Thus, early X-rays
play a major role in the evolution of the IGM temperature, 
as has been
emphasized earlier by Madau et al.\ (1997), Tozzi et al.\ (2000), and Chen
\& Miralda-Escud\'e (2003).

In the third simulation we have excluded the shock heating 
as well (in addition to Lyman-alpha heating and X-ray heating). In other
words, we only allowed photoionization heating as the single source of
heating of the gas in that simulation. The result of that simulation
is shown as a long-dashed
line in Fig.\ \ref{figME}. This change makes a dramatic difference - most
of the gas heating at high redshifts is due to shocks (i.e.\ structure
formation). However, the cosmological parameters are thought to be
sufficiently well known (Tegmark et al.\ 2004) that uncertainties in the
shock heating rate are not dominant. 

To further illustrate the role of X-rays in controlling the
$21\dim{cm}$ emission at high redshifts, we show with the solid grey line a
simulation in which all sources of ionization have a hard, quasar-like
spectrum, rather than a stellar spectrum as in the fiducial model. In that
case there is almost no stage in which the the redshifted $21\dim{cm}$ line
appears in absorption, because the gas heating is extremely
efficient. Finally, the dashed grey line shows the same model with
secondary electrons included. The conclusion we draw from Fig.\ \ref{figME}
is that the details of the early absorption phase in the $21\dim{cm}$
emission and the transition from absorption to emission depend on details
of the early universe physics, and, therefore, can be used as a powerful
test of the first episode of galaxy formation. On the other hand, the
observed evolution of the mean transmitted flux in the hydrogen Lyman-alpha
line (White et al.\ 2003) strongly constrains the late stages of the
reionization process and the predicted $21\dim{cm}$ emission for redshifts
less than about 9. In the rest of this paper, we focus on this late
reionization stage when we need a robust estimate of the cosmological
$21\dim{cm}$ signal.

\begin{figure}[t]
\plotone{\figdir/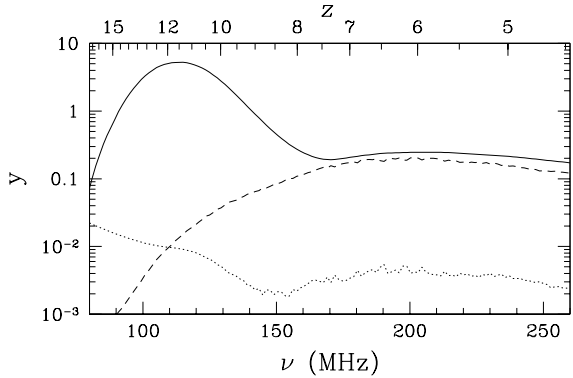}
\caption{\label{figYA}
  The evolution of the mass-weighted average 
  efficiency $y$ of coupling between the gas kinetic
  temperature and the spin temperate of the $21\dim{cm}$ line in the
  fiducial model A for three main coupling mechanism: Lyman-$\alpha$
  pumping (solid line), collisions with neutral atoms (dotted line), and
  collisions with free electrons (dashed line).
}
\end{figure}
Finally, in a further exploration of various physical effects responsible
for the redshifted $21\dim{cm}$ emission, we show in Figure \ref{figYA}, 
at the request of the referee, evolution of the mass-weighted average
coupling efficiency $y$, which is defined via the following equation: 
\begin{equation}
  T_S = {T_{\rm CMB}+y T_K\over 1+y},
\end{equation}
where $T_K$ is the gas kinetic temperature, and $T_S$ is the spin
temperature of the redshifted $21\dim{cm}$ line. Three independent physical
processes make contributions to $y$: pumping by Lyman-$\alpha$ photons, and 
collisions with neutral atoms and free electrons (Madau et al.\ 1997; Tozzi
et al.\ 2000), which are shown with different lines in Fig.\
\ref{figYA}. As one can see, the Lyman-$\alpha$ pumping dominates over the 
other two mechanisms for $z\ga8$, in agreement with Madau et al.\ (1997),
but even Lyman-$\alpha$ pumping becomes significant ($y>1$) only at
$z\la15$ (in our fiducial model), when shock heating already dominates by a
large margin.

An important conclusion of this demonstration is that the approximately
linear decrease in the brightness temperature with frequency in the range
$150\dim{MHz}\la\nu\la200\dim{MHz}$ is robust; it is a manifestation of the
reionization of the universe - the drop in the signal is simply due to
neutral hydrogen atoms becoming ionized.

\section{Mean Signal}

The first possible observable we consider is the mean (or global) emission
in the redshifted $21\dim{cm}$ line, averaged over the whole sky (as we
explain below, in practice this means signals on scales above about 2
degrees). 

\begin{figure}[t]
\plotone{\figdir/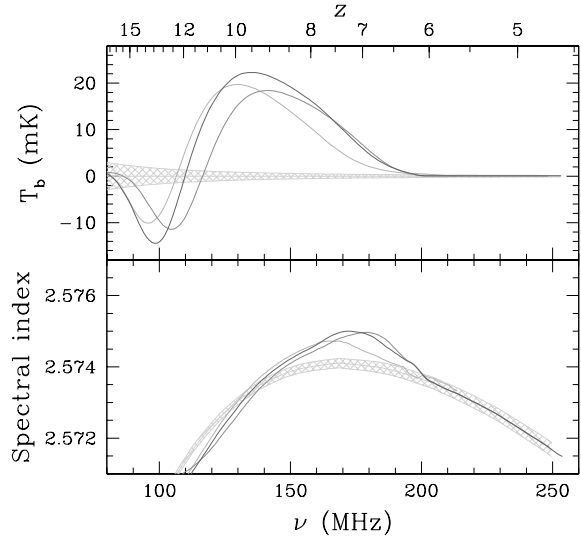}
\caption{\label{figMF}Top panel: The mean (over the whole sky) excess
  brightness temperature as a function of frequency and redshift for
  simulation sets A (dark gray), B (medium gray), and C (light gray). The
  thickness of the lightest gray (horizontal) band shows the $\pm5\sigma$
  sensitivity limit for a 10-day integration with a $5\dim{MHz}$
  bandwidth ($1\sigma=0.1\dim{mK}$ at $150\dim{MHz}$
  independent of the telescope size; see
  Shaver et al.\ 1999). We omit this limit in subsequent figures as it is
  so small. Bottom panel: Spectral index vs frequency for the three
  simulation sets with the galactic and extragalactic foreground included
  according to the model of Shaver et al.\ (1999). The lightest gray
  hatched band gives the $\pm5\sigma$ sensitivity limit for 10 days of
  integration (see explanation in the text).
}
\end{figure}
Figure \ref{figMF} shows the mean excess brightness temperature as a
function of frequency and redshift for the simulation sets A, B, and C,
and the resulting change in the spectral index of the total emission
(including the galactic and extra-galactic foreground). We used the model
of Shaver et al.\ (1999) for the foreground component in the bottom panel
of Fig.\ \ref{figMF}, which includes galactic thermal and non-thermal
components, and extragalactic sources. We also show an estimate of the
$\pm5\sigma$ sensitivity limit for a 10-day integration time for a system
temperature of $200\dim{K}$ (this limit is independent of telescope
size), as given by equation (6) of Shaver et al.\ (1999). 

The three theoretical curves in Fig.\ \ref{figMF} show the possible spread
in the predicted values when variations in cosmological parameters are
allowed. The predicted variation in the spectral index is well above the
sensitivity limits, and it is clear from both panels of Fig.\ \ref{figMF}
that sensitivity is not an issue in the detection of the $\HI$ signal. The
issue is how well the signal can be distinguished from the varying
foreground spectrum. (Calibration and radio interference are of course also
major concerns, and we address the reader to Shaver et al.\ (1999) for a
discussion of these technical matters.) 

The difference in the spectral index between the prediction for the total
signal (the cosmological signal plus foreground emission) is easily
distinguishable from the foreground, but, of course, we do not have the
privilege to observe the foreground signal alone. The foreground must be
estimated and the cosmological signal extracted from one observed total
signal. 
Shaver et al. (1999) considered various possible strategies, including
interpolations and extrapolations from frequencies outside the expected
signal range, and a "trend analysis" approach.
In the  present analysis we have tried fitting a low order polynomial to
the total signal as a surrogate for measuring the smooth foreground
signal, but we found the results to be marginal and strongly dependent
on the frequency range used to make a polynomial fit (this range is also
seriously constrained by the presence of FM bands below $108\dim{MHz}$
and TV bands above $174\dim{MHz}$). We have also tried to use a power
spectrum analysis of the spectral index to pick up high frequency
variations, and found this approach to be relatively insensitive to the
presence of the cosmological signal. Thus, we conclude that the mean
signal is not easily separable from the smooth foreground, although
sophisticated statistical approaches might yield more success.

As mentioned above, all simulations are normalized to reproduce
the observed mean transmitted flux in the spectra of SDSS quasars at
$z\sim6$. This observational constraint is a great advantage, as it  leaves
relatively small freedom in the possible variation of the excess brightness
temperature. However, there is uncertainty about the beginning and duration
of the pre-reionization era, which has been highlighted by the recent WMAP
results. 

\begin{figure}[t]
\plotone{\figdir/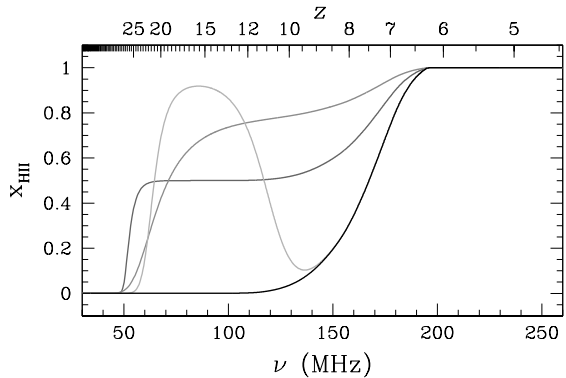}
\caption{\label{figXW}Ionization histories (volume averaged ionized
  fraction as a function of redshift) for our fiducial model A (black
  line; $\tau=0.06$) and three arbitrary models with $\tau=0.2$ (grey
lines).
}
\end{figure}

The first results from the WMAP experiment indicated a high value
($\tau\sim0.17$) of the optical depth to Thompson scattering (Kogut et al.\
2003). While these results have not been confirmed by a more comprehensive
analysis of the combined WMAP and SDSS data (Tegmark et al.\ 2004), the
possibility of a prolonged pre-reionization era remains an interesting one
(Madau, Ferrara, \& Rees 2001; Oh et al.\ 2001; Oh 2001; Venkatesan, 
Giroux, \& Shull 2001). In order to investigate the possible effect a
prolonged pre-reionization era might have on the observations of the
$21\dim{cm}$ emission, we have constructed three reionization models
that all produce an optical depth to Thompson scattering of $\tau=0.2$ -
about $1\sigma$ above the best-estimate value of Tegmark et al.\
(2004). These ionization histories are shown in Figure \ref{figXW} together
with out fiducial model from the simulation set A.

\begin{figure}[t]
\plotone{\figdir/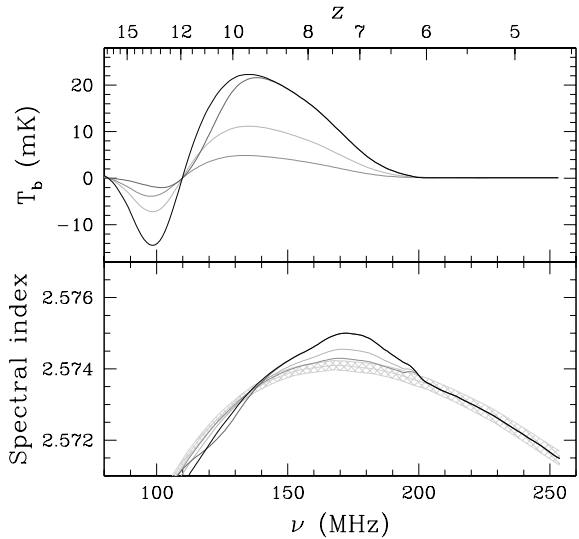}
\caption{\label{figMW}
The same as Fig.\ \ref{figMF}, but for
the ionization histories from Fig.\ \ref{figXW}.
}
\end{figure}
It is important to underline that these ionization histories are obtained
arbitrarily, and are not results of simulations - in fact, the nature of
ionizing sources that can produce such histories is not known. More than
that, the light grey curve, which is designed to mimic ``double
reionization'' (Cen  2003), is, in fact, unphysical, as low density
regions in the universe will not be able to recombine between redshifts 12
and 7. Hence, these ionization histories only serve to illustrate the
sensitivity of the redshifted $21\dim{cm}$ $\HI$ measurement as a tool
for discriminating various ionization histories. The respective predictions 
for the mean signal are shown in Figure \ref{figMW}. As one can see, the
$21\dim{cm}$ measurement may be used to constrain ionization histories
in which the universe remains partially ionized for a prolonged period of
time, but is not sensitive to the models in which the universe was reionized
twice. However, as we have mentioned above, such models are not very
plausible.

\section{Fluctuations in $21\dim{cm}$ Emission}

The $21\dim{cm}$ signal is not, of course, completely uniform over the
sky: there are several sources of fluctuations in the $21\dim{cm}$
signal at high redshift.

At the redshifts of interest ($z\sim10$), a given comoving size $\Delta x$
corresponds to the angular size 
$$
        \Delta\theta = 0^\prime.6 C_z {\Delta x\over h^{-1}\dim{Mpc}}
$$
and the range of frequencies in the redshifted $21\dim{cm}$ line
$$
        \Delta\nu = 0.1\dim{MHz} D_z {\Delta x\over h^{-1}\dim{Mpc}},
$$
where $C_z$ and $D_z$ are weak functions of $z$ in the redshift interval
$6<z<10$. For $\Omega_{m,0}=0.35$ at $z=9$, $C_z=0.938$ and $D_z=0.894$. 
Thus, any realistic observation with $\Delta\nu\ga1\dim{MHz}$ and
$\Delta\theta\ga1^\prime$ will probe the linear regime at $z\sim7-12$
with $\Delta x\ga2h^{-1}\dim{Mpc}$.

The linear large-scale density fluctuations are the subject of this
paper. As mentioned above, another possible source of fluctuations in the
high redshift $21\dim{cm}$ emission are $\HII$ regions created by bright
quasars well before the moment of reionization (Madau et al. 1997; Tozzi et
al.\ 2000). Such $\HII$ regions, if big enough, could be prominent features
scattered sparsely over the sky. However, the simulations presented in this
paper do not encompass sufficiently large computational volumes to include
luminous quasars. Models that include a representative sample of bright
quasars require a different computational approach, and will be the subject
of a future paper. 

Fluctuations can also be caused by variations in the quantity 
$q=1-T_{\rm CMB}/T_S$, where $T_S$ is the gas spin temperature. However,
the simulations that we use in this paper suggest that such variations are
negligible compared to the density variations.

Given the power spectrum of density fluctuations $P(k)$ (hereafter we
assume that the gas follows the dark matter on large scales), the rms
density fluctuation as a function of the angular resolution $\Delta\theta$
and the frequency bandwidth $\Delta\nu$ is given by the following equation:
\begin{equation}
        \delta^2_{\rm RMS} =
        {1\over(2\pi)^3}\int\limits^\infty_{-\infty}dk_\parallel
        \int d^2k_\perp
        P\left(k\right)\,
        e^{-k_\parallel^2R_\parallel^2-{\vec{k}}_\perp^2R_\perp^2},
        \label{delrms}
\end{equation}
where $k_\parallel$ is the wavenumber along the line-of-sight (the
frequency dimension), $\vec{k}_\perp$ is the wavevector in the plane of
the sky (two angular dimensions), $k\equiv
\sqrt{k_\parallel^2+{\vec{k_\perp}}^2}$, and
$$
        R_\parallel = 1h^{-1}\dim{Mpc} {\Delta\nu\over 2.35\times
	  0.1\dim{MHz}D_z}
$$
and
$$
        R_\perp = 1h^{-1}\dim{Mpc} {\Delta\theta\over 2.35\times
	  0^\prime.6 C_z},
$$
and the factor $2.35=2\sqrt{2\log2}$ comes from the conversion from the
FWHM to the gaussian dispersion.

\begin{figure}[t]
\plotone{\figdir/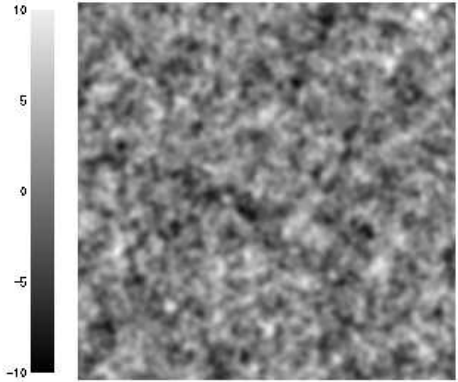}
\caption{\label{figIM}A $20^o\times20^o$ image of fluctuations in the
  excess brightness temperature of the redshifted $21\dim{cm}$ line at $z=9$
  for our fiducial model (with $4^\prime$ resolution and $1\dim{MHz}$
  bandwidth). The color-bar on the left shows the scale in $\dim{mK}$.
}
\end{figure}
As an example, Figure \ref{figIM} shows linear fluctuations in a
$20^o\times20^o$ region of the sky. Because the fluctuations are gaussian,
there are no distinct features in the image, and separating this signal
from the random fluctuations in the foregrounds is a non-trivial task. A
big enough quasar will create an $\HII$ region around itself, and will form
a significant structure in the fluctuation map (Madau et al.\ 1997) - as
mentioned above, this effect will be considered in a follow-up paper. 

\begin{figure}[t]
\plotone{\figdir/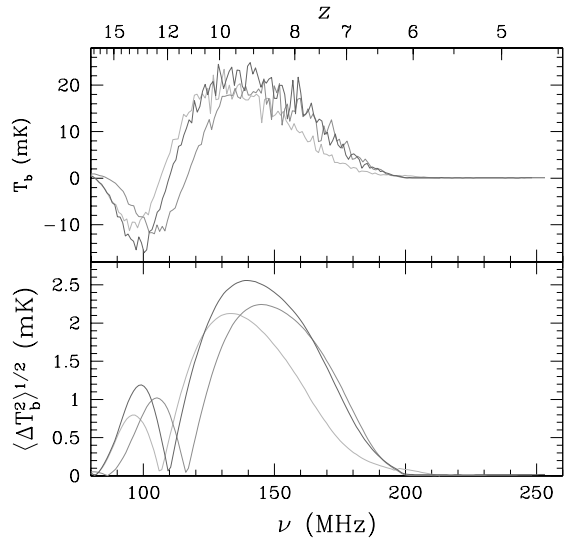}
\caption{\label{figFF}The excess brightness temperature (top) and its rms
  fluctuation (bottom) measured in a beam of $4^\prime$ FWHM with
  $1\dim{MHz}$ bandwidth.
}
\end{figure}
A complimentary view of the fluctuations in the redshifted $21\dim{cm}$
emission is provided in Figure \ref{figFF}, which shows the excess
brightness temperature and its rms fluctuations for our three simulation
sets as a function of frequency. The level of rms fluctuations on the scale
we consider is only about 10\% of the mean signal. 

In order to determine whether the fluctuations are potentially observable,
they have to be compared with the possible fluctuations in the foreground
(galactic and extragalactic) emission.

\subsection{Angular Fluctuations}

\begin{figure}[t]
\plotone{\figdir/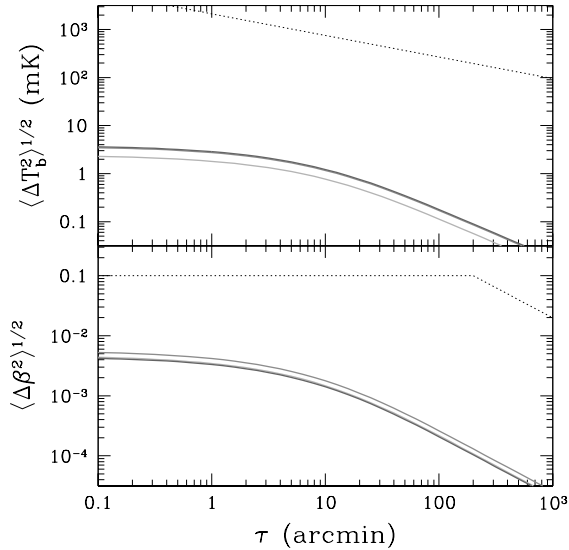}
\caption{\label{figFA}The rms excess brightness temperature fluctuations in
  $1\dim{MHz}$ bandwidth at $z=8.5$ (top) and the rms {\it angular\/}
  fluctuation in the spectral index $\beta$ (bottom) as a function of the
  angular scale. The dotted lines show the foreground contaminations:
  fluctuations due to residual point sources after all sources above
  $1\mu\dim{Jy}$ have been removed (top; Di Matteo et al.\ 2002) and
  variations in the spectral index of the galactic synchrotron emission
  (bottom; Banday \& Wolfendale 1991). See the discussion in text for full
  explanation. 
}
\end{figure}
As has been discussed by Di Matteo et al.\ (2002), angular fluctuations
due to extragalactic radio sources are significantly larger than those
from the pre-reionization era. In order to estimate the contribution of the
extragalactic sources, Di Matteo et al.\ (2002) adopted a model for
differential source counts at $150\dim{MHz}$ at low flux density levels,
$N(S) \approx 700\dim{sr}^{-1}\dim{mJy}^{-1} (S/1\dim{mJy})^{-1.75}$,
and a model for the source correlation function, 
$w(\theta)\approx(\theta/\theta_0)^{-0.85}$ with
$\theta_0=4^\prime$. While the indexes of the power-laws adopted by Di
Matteo et al.\ (2002) agree with other studies, there exist surprising
differences in the values of the amplitudes they adopted. For example,
Hopkins et al. (1998) give the source number counts as
$N(S) \approx 10\dim{sr}^{-1}\dim{mJy}^{-1} (S/1\dim{mJy})^{-1.75}$ at
$1.4\dim{GHz}$. Assuming that no new source population is present at
$150\dim{MHz}$ that is not observed at $1.4\dim{GHz}$, and that the
source spectral index between $150\dim{MHz}$ and $1.4\dim{GHz}$ is about
-0.75, the source flux density of $1\dim{mJy}$ at $1.4\dim{GHz}$ translates
into $5.4\dim{mJy}$ at $150\dim{MHz}$, which gives the source
number counts at $150\dim{MHz}$ at $S=1\dim{mJy}$ of about
$200\dim{sr}^{-1}\dim{mJy}^{-1}$, a factor of 3.5 below the value
adopted by Di Matteo et al.\ (2002). In addition, both FIRST and NVSS
measurements of the extragalactic radio source correlation function give
the functional form adopted by Di Matteo et al.\ (2002) but with
$\theta_0\approx1^\prime$ (Blake \& Wall, 2002). The latter applies for
sources brighter than a few mJy at $1.4\dim{GHz}$ - significantly brighter
than sources considered by Di Matteo et al.\ (2002) - but it is difficult
to imagine that faint sources are clustered more strongly than the bright
ones. The total effect of these corrections would be to reduce the Di
Matteo et al.\ (2002) estimate of the foreground fluctuations by a factor
of 11. 

On the other hand, Di Matteo et al.\ (2002), as one of their cases, showed
their estimate when all sources above $1\mu\dim{Jy}$ have been removed from
the data (thick black line in Fig.\ 1 of Di Matteo et al.\ 2002), which is,
probably, an overly optimistic assumption. If we conservatively assume that
only sources above $1\dim{mJy}$ can be effectively removed from the data or
adequately modeled in the data analysis, then the estimate of angular
fluctuations from the extragalactic radio sources would be a factor of 10
higher - but that is almost exactly compensated by the factor of 11 by
which Di Matteo et al.\ (2002) overestimated the clustering signal of
extragalactic point sources. Therefore, for the rest of this paper, we
adopt the thick solid line from Fig.\ 1 of Di Matteo et al.\ (2002) as a
reasonable estimate for the angular fluctuations in the $21\dim{cm}$ signal
due to the clustering of extragalactic radio sources.

As can be seen from the top panel of Figure \ref{figFA}, our results fully
agree with the conclusions of Di Matteo et al.\ (2002): the contamination
by the extragalactic source population makes it highly unlikely that the
predicted levels of the $21\dim{cm}$ fluctuation signal from the
pre-reionization era can be reached.

Not only the extragalactic radio sources, but also the even stronger
galactic foreground emission will produce substantial
angular fluctuations in the emission over the sky. Such fluctuations may
extend down to quite small angular scales, and can only add to the
already dominant fluctuations from the extragalactic source population.
The small contribution from galactic thermal emission will certainly add
to the fluctuation signal - it may be similar to the structure in the
galactic infrared cirrus. There may in addition be other contributors to
the foreground angular fluctuation signal, such as the extended radio
emission from clusters of galaxies. Thus, in view of the overwhelming
foreground contamination, it seems highly unlikely that the angular
fluctuations in the redshifted $\HI$ signal will by themselves be useful
in studying the pre-reionization era.

Can one possibly use frequency information to help separate the angular
fluctuations in the HI signal from the foreground fluctuations, as the
spectra of the foregrounds are expected to be smooth, with only gradual
changes in spectral index? In order to investigate this possibility, we
computed the angular variation in the spectral index on the sky as the
difference between two images taken $5\dim{MHz}$ apart. The bottom
panel of Fig.\ \ref{figFA} shows the rms spectral index fluctuations as a
function of angular scale at $150\dim{MHz}$ ($z=8.5$). Unfortunately,
spatial variations in the spectral index of the galactic synchrotron
emission make this approach difficult if not impossible. Banday \&
Wolfendale (1991) estimate variations of the order of 0.1 on angular
scales of $3^o\!.5$. While their estimate applies to the frequency range
from $400$ to $800\dim{MHz}$, it is reasonable to assume that similar
variations exist at frequencies around $150\dim{MHz}$. The dotted line
in the bottom panel of Fig.\ \ref{figFA} shows a simple toy model for
the variation in the spectral index of the galactic synchrotron
emission: we take Banday \& Wolfendale's (1991) value of  0.1 at
$3^o\!.5$, and then assume that fluctuations are uncorrelated on larger
scales and remain fixed on smaller scales. This lower limit is overly
conservative, since Banday \& Wolfendale (1991) indicate
that fluctuations in the spectral index are correlated on angular scales
above $3^o$, and on smaller angular scales fluctuations are expected to
be higher because the angular scale of $3^o$ does not correspond to any
particular physical scale.

There may be other more sophisticated techniques that can be employed in
analyzing the observed data cube. However, on the basis of  present
knowledge, we conclude that the angular fluctuations will be extremely
difficult if not impossible to utilize in measuring and studying the $\HI$
signal.

\subsection{Fluctuations in the Frequency Domain}

If the angular fluctuations are not observable, one may still hope to use
the fact that all known radio foregrounds are expected to be relatively
smooth functions of frequency, and rely on the frequency information alone
to separate the cosmological signal. For the cosmological signal the
frequency direction is just another spatial dimension, and so the
cosmological signal should fluctuate in frequency space very much as it
fluctuates over the sky. By contrast, the foreground contamination is
totally different in the angular and frequency dimensions: the foreground
angular signal has all the structure of a distribution of point sources
over the sky, whereas the foreground frequency signal is just a sum of
smooth spectra, hence itself a smooth function, against which the
relatively sharp cosmological fluctuations (in frequency) should stand out
clearly. This approach may in fact be by far the best one to studying the
reionization $\HI$ signal. Other authors (e.g.\ Di Matteo et al.\ 2002)
have also recognized this possibility; here we quantify it and show the
dependence on observational parameters such as beam size and bandwidth.

\begin{figure}[t]
\plotone{\figdir/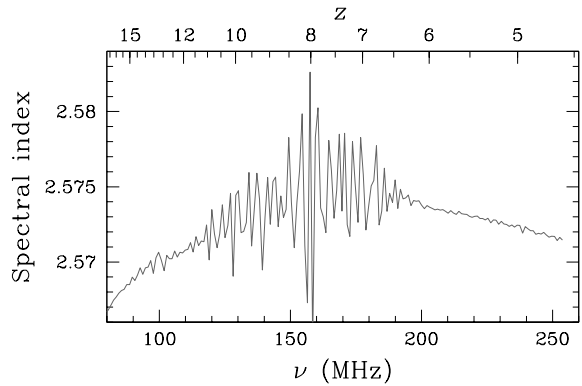}
\caption{\label{figFS}The spectral index of the total signal ($\HI$ signal
  plus the foregrounds) in a beam of $4^\prime$ FWHM with a
  $1\dim{MHz}$ bandwidth.
}
\end{figure}
\begin{figure}
\plotone{\figdir/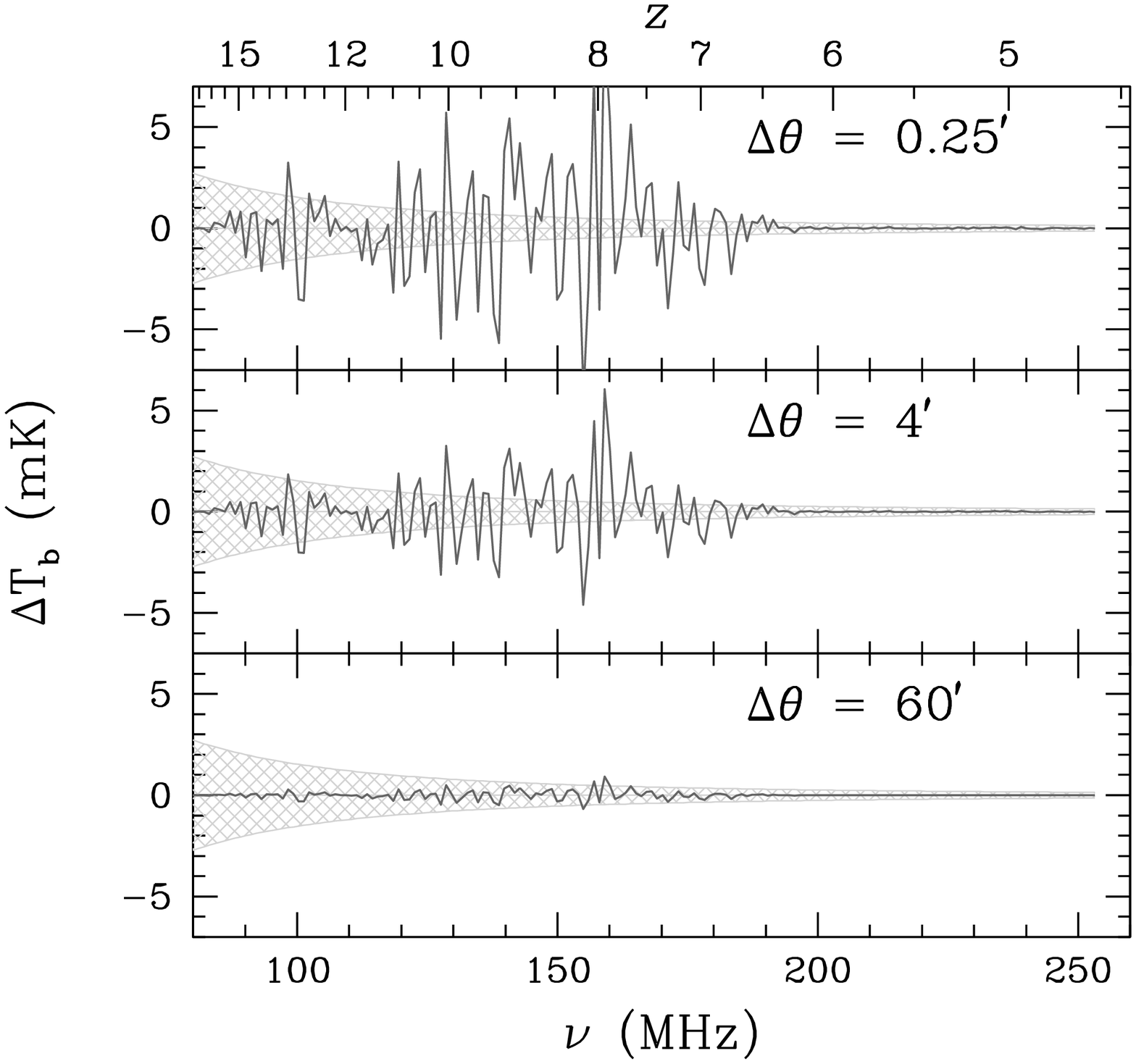}
\plotone{\figdir/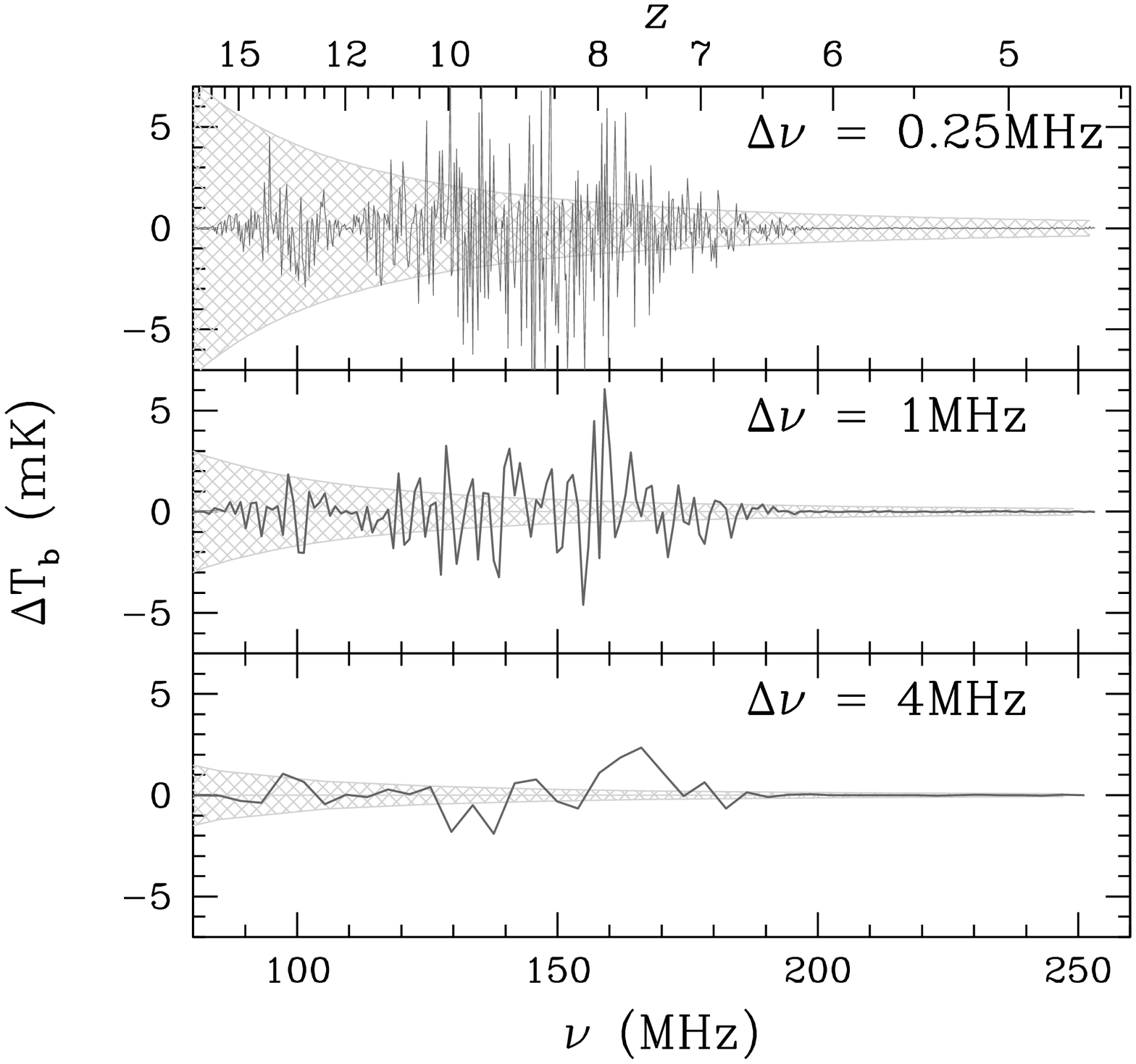}
\caption{\label{figFU}(a) The excess brightness temperature fluctuations
  (with the mean signal substructed) for an observation with a fixed
  bandwidth of $1\dim{MHz}$ and varied beam size at $150\dim{MHz}$
  (assuming an appropriate telescope size). The hatched regions give
  $\pm5\sigma$ sensitivity limit for a 40-day integration time for the
  foreground model of Shaver et al.\ (1999). (b) The same as 
  (a) but for a fixed beam size of $4^\prime$ at $150\dim{MHz}$ and varied
  bandwidth.
}
\end{figure}
Figure \ref{figFS} 
shows the $\HI$ fluctuations superimposed on the combined spectrum
of the galactic and extragalactic foregrounds (spectral index is used
here in order to be able to show both the $\HI$ and foreground signals on a
single plot). The difference between the frequency structure of the $\HI$
signal and the foregrounds is immediately obvious, and it is clear that
one may study the former with not too much complication from the latter.
Thus, it appears that the best strategy for measuring the reionization
signal may be to forsake any attempt to measure angular fluctuations and
concentrate on frequency fluctuations alone. Thanks to the relative
spectral smoothness of the foreground emissions, this is almost certainly the
best approach to detect and study the reionization $\HI$ signal.
(Note that we assume here that the beamsize is independent of
frequency; the effect of a frequency-dependent beam is considered
below.)

The excess brightness temperature fluctuations (with the mean signal
subtracted) are shown in Figure \ref{figFU} 
for several values of the beam size
and bandwidth. As can be seen from the figures, the frequency
fluctuations are a strong function of the beam size (which we specify at
$150\dim{MHz}$) and the bandwidth, but for a large range of beam sizes
and bandwidths they are well above the sensitivity limits (for example,
for a $4^\prime$ beam and $1\dim{MHz}$ bandwidth the signal-to-noise
ratio at $150\dim{MHz}$ is about 22 for a 40-day integration).

\begin{figure}[t]
\plotone{\figdir/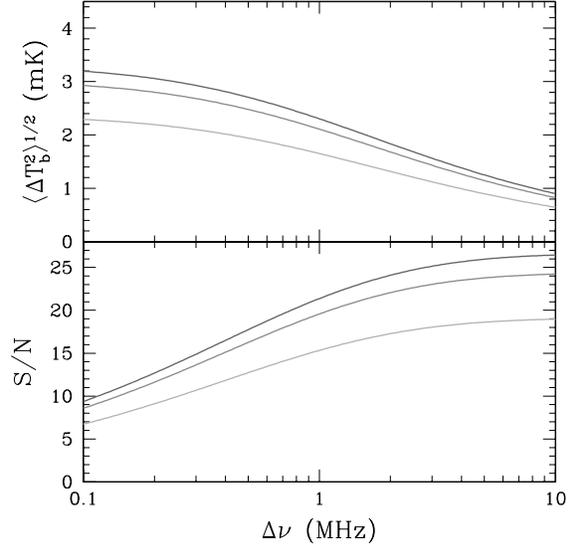}
\caption{\label{figFN}The rms excess brightness temperature fluctuations on
  angular scales of $4^\prime$ at $z=8.5$ as a function of bandwidth (top) 
  and the respective signal-to-noise ratio (bottom) for a 40-day
  integration.
}
\end{figure}
The fluctuations in the frequency domain depend on the bandwidth. Is
there an optimal range for the value of the bandwidth? Obviously,
fluctuations are smaller on the larger scales. However, the sensitivity
is also higher for larger bandwidths. It turns out that these two
effects almost cancel each other: Figure \ref{figFN} shows that for
bandwidths in excess of about $1\dim{MHz}$, the signal-to-noise ratio is
essentially constant, so the exact value of the bandwidth is probably
not that important. 

\begin{figure}
\plotone{\figdir/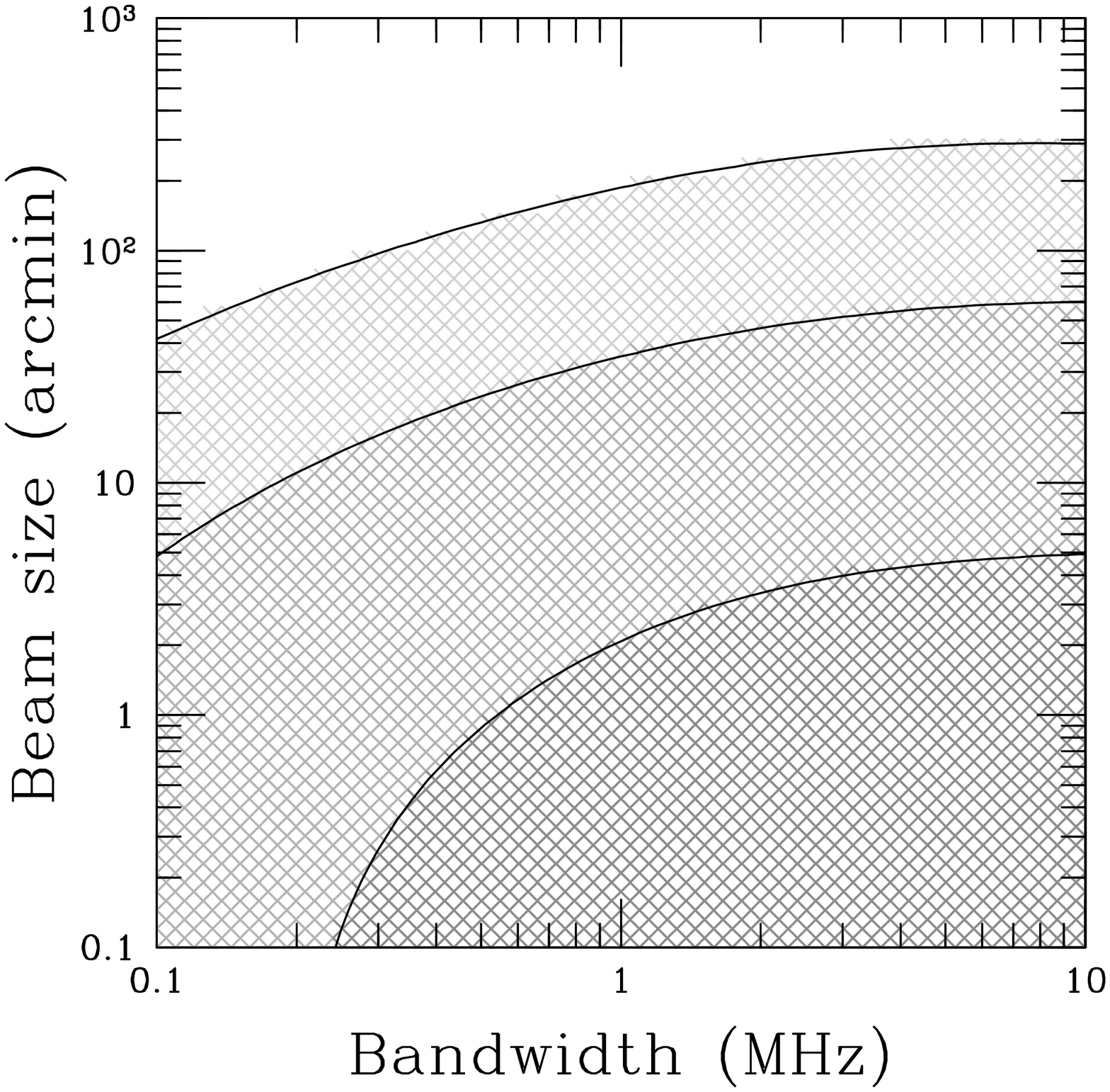}
\plotone{\figdir/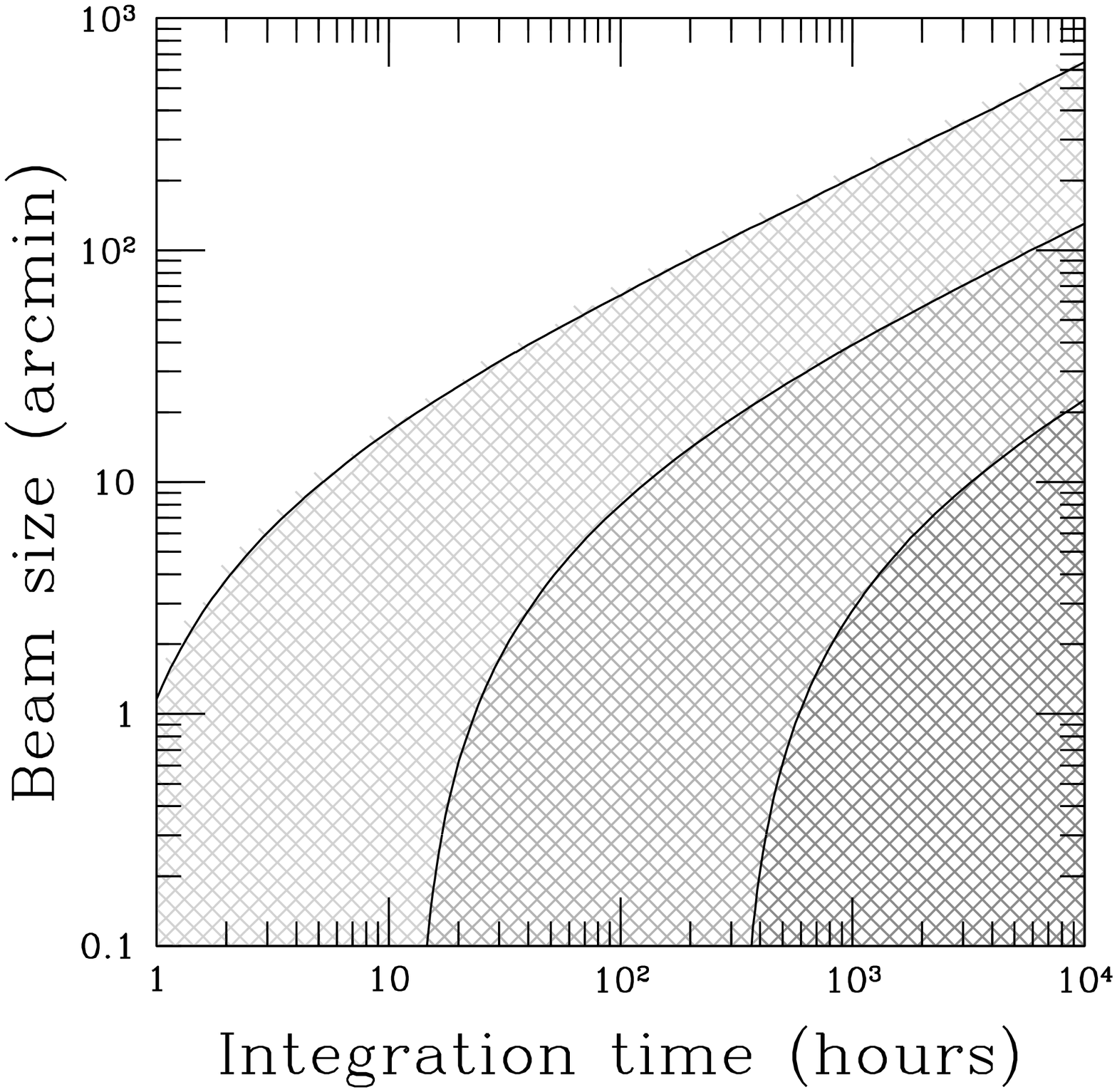}
\caption{\label{figOB}Contours of signal-to-noise ratio in the fiducial
  model for the rms excess brightness temperature fluctuation at
  $150\dim{MHz}$ ($z=8.5$) and assuming the system temperature of
  $200\dim{K}$ as a function of bandwidth and {\it filled aperture\/} beam
  size for the 
  integration time of 40 days (a) and as a function of integration 
  time and the beam size for $1\dim{MHz}$ bandwidth (b). Shown are
  contours of S/N=1 (light gray area), S/N=5 (medium gray area), and S/N=25
  (dark gray area). The signal-to-noise will be respectively lower for
  extended reionization models.
}
\end{figure}
In order to illustrate the measurability of the fluctuation signal in
the frequency domain, we present in Figure \ref{figOB} contours of
fixed signal-to-noise ratio for various beam sizes, bandwidths, and
integration times, as a guide for developing an observational
strategy. Since our fiducial model is based only on known physical
processes and does not include an extended period of reionization, it
can serve as an upper limit for the expected signal (except for the
occasional presence of a luminous QSO in the beam). Thus, the
signal-to-noise ratio of a real experiment may be smaller, but generally
will not be larger, than the one shown in Fig.\ \ref{figOB}.

The top panel of Fig.\ \ref{figOB} also shows what angular and frequency
scales one needs in order to measure the mean signal: S/N=3 for the
fluctuations essentially means that fluctuations are unobservable, so
different patches of the sky larger than about 2 degrees will have the
same signal within the sensitivity limit (for a 40-day integration).

The feasibility of measuring the frequency signal will, of course,
critically depend on the ability to calibrate different frequency bands
to within a fraction of a $\dim{mK}$. Such an observation
may be done, for example, by beam
switching between the target area on the sky and a strong nearby
extragalactic radio source, whose spectrum is expected to be very smooth
on $1\dim{MHz}$ frequency scales. This approach
would be difficult or impossible
for the detection of the mean (all-sky) signal, as discussed by Shaver
et al. (1999), because the frequency dependence of the mean $\HI$ signal is
relatively smooth, but in the case of the sharp frequency variations of
the fluctuation signal it should become possible. This could be a great
advantage because relative measurements are far easier to make than
absolute measurements.

There will be some natural narrow-band contaminants of the frequency
signal, such as galactic radio recombination lines. However these show
up at well-known frequencies (separated by typically $1-2\dim{MHz}$
 in this
frequency range), and they are narrow, so they can be removed from the
data. A major potential contaminant, of course, is man-made RFI.
However, it is variable and generally very narrow-band; techniques are
being developed to identify and excise such contamination. How
completely such contamination can be removed remains to be determined,
however. To the extent that spectral contaminants show up predominantly
in emission, whereas the fluctuations in Fig.\ \ref{figFU} 
go both ways, in emission and in absorption relative to the
smooth foreground, it may be possible to use these properties to help
identify and characterize the signal. Furthermore, as the cosmological
parameters to be extracted from the gaussian $\HI$ signal are the same in
all directions of the sky, observations in different directions 
should ultimately give the same results.

In addition to the contaminants mentioned above, there is another
effect that can potentially obscure the cosmological $\HI$ signal: the
leakage of the angular fluctuations into the frequency domain due to the
frequency-dependent beam size (Oh \& Mack 2004; Di Matteo et al.\
2002). 
In the simplest case of a single filled
aperture, it is inversely proportional to the frequency of observations. In
order to see fluctuations in the cosmological $\HI$ signal, one has to
observe over a range of frequencies, and because observations at different
frequencies have different beam sizes, they sample different regions of the
sky. The angular fluctuations between the observed brightness temperature
in these different regions appear as fluctuations in the frequency domain.

It is straightforward to estimate this effect. Using equation (4) of Di
Matteo et al.\ (2002), we approximate the power spectrum of fluctuations
in a beam with a gaussian width $\sigma$ as
\begin{equation}
  C_l = A^2 l^{\beta-2} e^{-l^2\sigma^2/2},
\end{equation}
where $C_l$ is the total power in a spherical multipole $l$,
$A$ is the amplitude of fluctuations at $l=1$, and $\beta\approx0.85$
is the power-law index of the fluctuation angular correlation function.

Let $T(\vec{\Omega},\sigma)$ be a temperature fluctuation in a beam
with a gaussian width $\sigma$ in the direction $\vec{\Omega}$ on the
sky. Two observations in the direction $\vec{\Omega}$ spaced in
frequency by $\Delta\nu$ would see the rms difference in the brightness
temperature:
$$
\delta T_{\rm RMS} \equiv \Delta\sigma \left\langle\left(
{dT(\vec{\Omega},\sigma)\over d\sigma}\right)^2\right\rangle^{1/2} =
$$
\begin{equation}
\phantom{AAA}
\sqrt{\beta(\beta+1)} {\Delta\sigma\over\sigma} T_{\rm RMS},
\label{dtrms}
\end{equation}
where $T_{\rm RMS}$ is the rms angular temperature fluctuation in the
beam
$\sigma$, and $\Delta\sigma/\sigma=\Delta\nu/\nu$.
For a beam with FWHM of $4^\prime$ this estimate gives a value
of about $6\dim{mK}$ at $150\dim{MHz}$ for $T_{\rm RMS}=1\dim{K}$.
This fact was noted by Di
Matteo et al.\ (2002) and Oh \& Mack (2004) as a severe obstacle for
measuring cosmological fluctuations in the frequency domain. However,
these previous works missed the fact that the leaked angular fluctuations are
strongly correlated. Let us consider three beams A, B, and C spaced by
$\Delta\nu$ in frequency, with beam sizes $\theta_A$, $\theta_B$, and
$\theta_C$. The signal in beam A will differ from the signal in beam B by
some amount due to sources located within the ring on the sky with a radius
of $(\theta_A+\theta_B)/2$ and the width of
$\theta_B-\theta_A$. Analogously, the signal in beam B will differ from the
signal in beam C due to sources located within a ring from $\theta_B$ to
$\theta_C$. But because these two rings are adjacent to each other, sources
within the ring $\theta_B-\theta_A$ are strongly correlated with the
sources within the ring $\theta_C-\theta_A$, so that the increase or
decrease in the total signal in beams B and C due to fluctuations will be a
relatively smooth function of frequency as well, and the frequency
fluctuations from the leakage signal will be much smaller. Namely, given a
temperature difference $T_{BA}$ between beams B and A and the temperature
difference $T_{CB}$ between beams C and B, the rms difference between these
two temperatures  $\Delta T_{\rm RMS} \equiv \left\langle\left(T_{CB} -
T_{BA}\right)^2\right\rangle^{1/2}$ is 
$$
\Delta T_{\rm RMS} = \Delta\sigma \left\langle\left(
{d \delta T(\vec{\Omega},\sigma)\over
d\sigma}\right)^2\right\rangle^{1/2}
=
$$
\begin{equation}
\sqrt{2\beta(\beta^2+4\beta+7)} \left(\Delta\sigma\over\sigma\right)^2
 T_{\rm RMS},
\label{ddtrms}
\end{equation}
another factor of $\Delta\nu/\nu\approx 1/150$ smaller.

\begin{figure}[t]
\plotone{\figdir/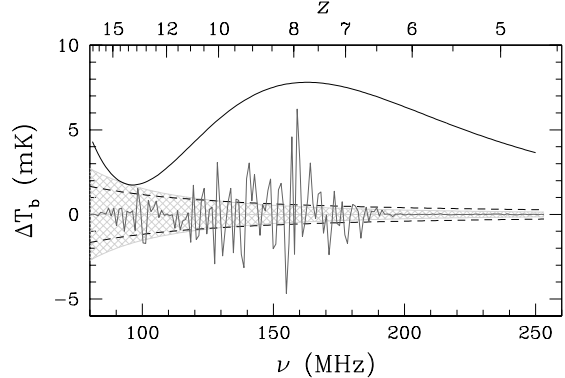}
\caption{\label{figBL}The excess brightness temperature fluctuations
  (with the mean signal substructed) for an observation with a fixed
  bandwidth of $1\dim{MHz}$ and a {\it frequency-dependent\/}
   beam size of $4^\prime$ at $150\dim{MHz}$ (similar to the middle panel
  of Fig.\ \ref{figFU} but with the frequency dependence of the beam size taken
  into account) together with a realization of a
  frequency fluctuation due to leakage of angular fluctuations into the
  frequency domain (solid black line). The hatched regions give
  $\pm5\sigma$ sensitivity limit for a 40-day integration time for the
  foreground model of Shaver et al.\ (1999), and the black dashed lines
  show characteristic $\pm5\sigma$ errors due to angular fluctuation
  leakage (eq.\ [\ref{ddtrms}]).
}
\end{figure}
In order to illustrate this fact further, we have simulated the leakage
effect by constructing a sample of the sky with gaussian 
fluctuations\footnote{The assumption of gaussianity is a good one as long
  as the number of point sources within the beam is large, which is the
  case for the values of the beam size considered here.}
having a power-law power spectrum, as given by equation (4) of Di
Matteo et al.\ (2002). We then computed the temperature within the gaussian
beam of variable size $\theta = 4^\prime(\nu/150\dim{MHz})$
for $\nu$ ranging from 80 to $250\dim{MHz}$ in increments of
$1\dim{MHz}$ (the bandwidth used in the middle panel of Fig.\ \ref{figFU}),
centered at a random point on
the simulated region of the sky. One realization of the leakage signal
within the beam is shown in Figure \ref{figBL}.
The black solid line gives the leaking angular fluctuation $\delta T$,
and the two black dashed lines
show the $\pm5\sigma$ levels of contamination of the cosmological $\HI$
signal expected in this case ($5\times\Delta T_{\rm RMS}$).
As one can see, the contamination is comparable to the
observational noise for a 40-day integration, so it becomes the
dominant component of noise for longer integration times. Thus,
it will not be useful to use integration times well in excess of 40
days on a given field.

\begin{figure}
\plotone{\figdir/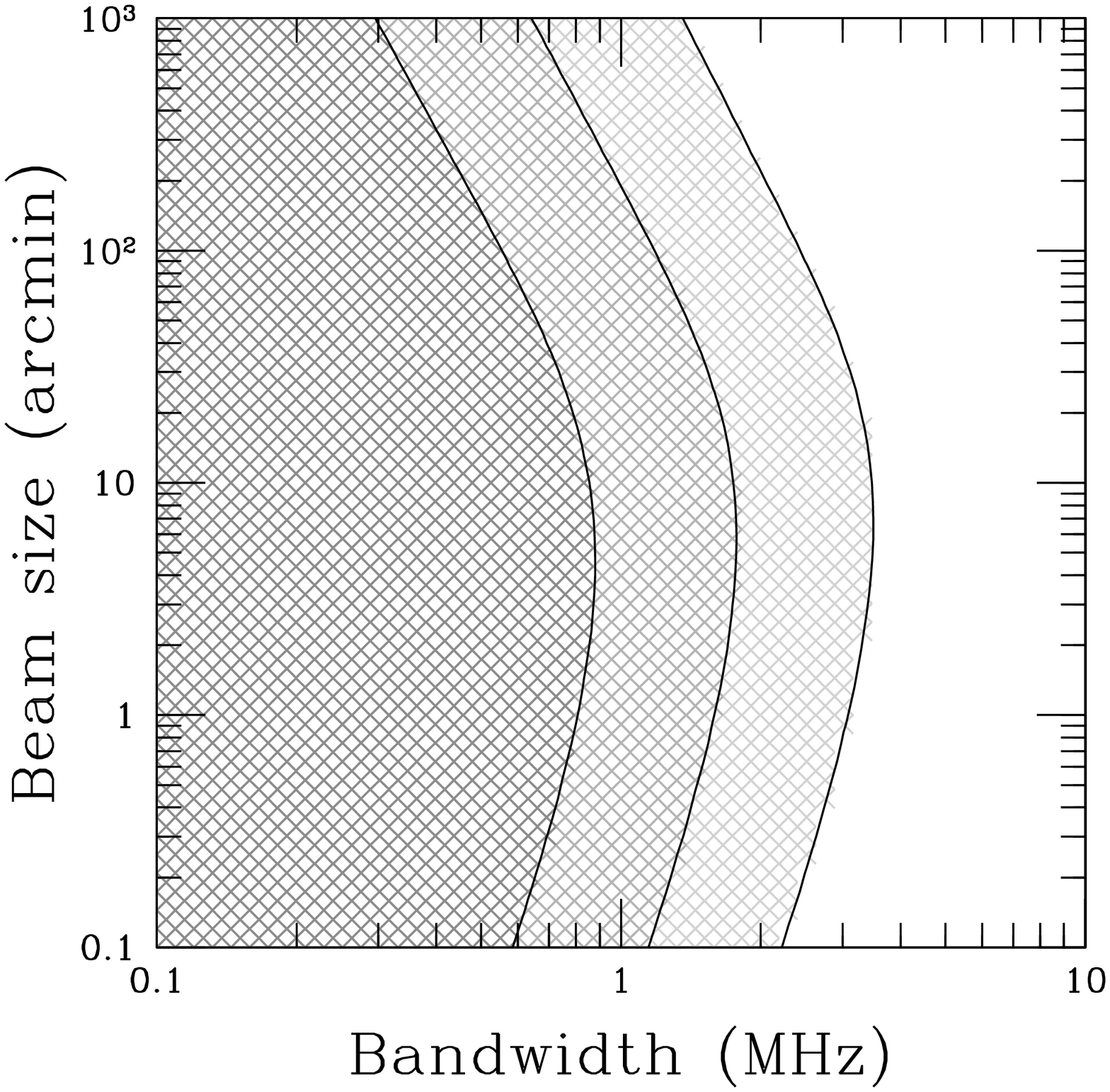}
\plotone{\figdir/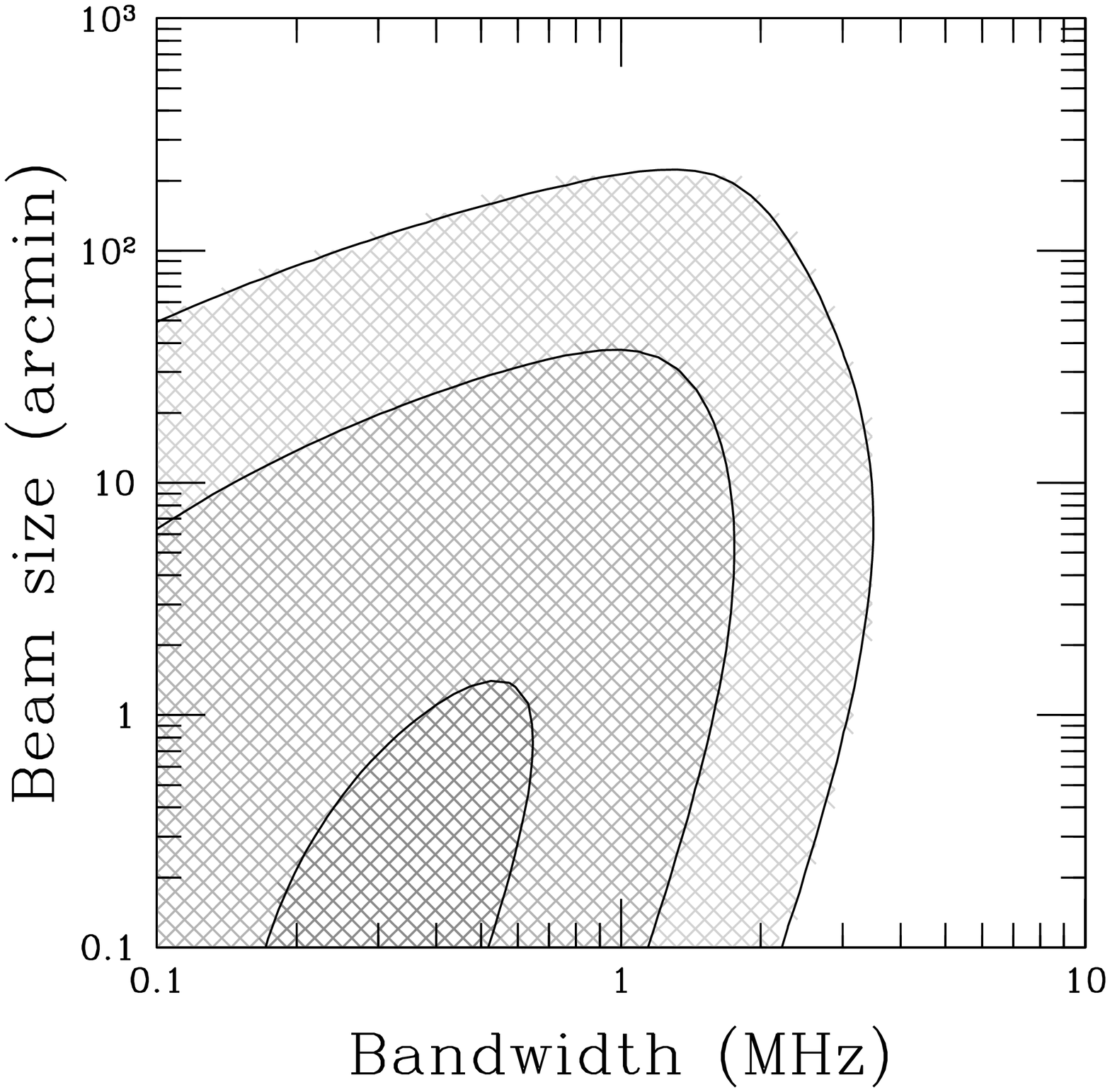}
\caption{\label{figOC}(a) Contours of signal-to-noise ratio in the fiducial
  model for the rms excess brightness temperature fluctuation at
  $150\dim{MHz}$ ($z=8.5$) and assuming a system temperature of
  $200\dim{K}$ as a function of bandwidth and {\it filled aperture\/} beam
  size with the noise entirely due to angular fluctuation leakage
  (eq. [\ref{ddtrms}]). Shown are  contours of S/N=1 (light gray area),
  S/N=5 (medium gray area), and S/N=25 (dark gray area). (b) Contours
  of signal-to-noise ratio for the same observational parameters, with both
  observational noise (for a 40-day integration) and the leakage noise,
  included in quadrature. 
}
\end{figure}
Figure \ref{figOC} illustrates possible observational choices when the
angular fluctuation leakage is included as a source of noise. The top panel
of Fig.\ \ref{figOC} shows contours of signal-to-noise ratio with only
leakage noise included - these contours scale in inverse proportion to the
amplitude of angular fluctuation on the beam scale. The bottom panel
includes both the observational noise and the leakage noise. While the
bottom panel is the most complete representation of possible observational
signal-to-noise ratio for our model, the contours on the bottom panel 
depend on both the integration time and the amplitude of angular
fluctuations on the beam scale and scale non-trivially when these
quantities are changed. Therefore, while this panel can be used as an
illustration of possibilities, it is only applicable for the values of
observational or theoretical parameters adopted in this paper.

Other sources of angular fluctuations that might produce leakage in the
frequency domain are the galactic synchrotron and thermal emission.  If
the galactic emission is not structured more strongly on scales of tens
of arcminutes than the extragalactic point source contamination, then it
will not present a significant obstacle to measuring the cosmological
$\HI$ signal, but this question requires further investigation,
including more detailed observations of the galactic emission.

There are various ways in which the leakage of angular fluctuations
could be minimized. The most important would obviously be to observe
with the same beam size at all frequencies. This could best be
approximated with the use of scaled arrays especially designed to control
the synthesized beam and sidelobes. For filled-aperture telescopes,
frequency-dependent illumination of the reflector would be required, at
least over a limited frequency range, but this would then utilize only a
fraction of the total aperture, and would undoubtedly be a major
technical challenge. In any case it would clearly be advantageous to
observe in directions of the sky where the foreground contamination is
lowest: regions of low extragalactic source density and fluctuations,
and regions of high galactic latitude where the galactic emission is
weakest and smoothest.

\section{Conclusions}

The challenge in observing the redshifted $21\dim{cm}$ line of neutral
hydrogen from the pre-reionization era is not that it is weak, but
rather that it is hidden behind strong foreground
emission (both galactic and extragalactic). Our results confirm the
conclusions of other authors that detection of the angular fluctuations
on the sky will be extremely difficult or impossible because of the huge
contamination from fluctuations in the foreground emissions.

Better opportunities are provided by the fact that the radio foregrounds
are slowly varying as a function of frequency. This generic feature
might make it feasible (although very challenging) to detect the overall
mean signal from the pre-reionization era. This signal would be the true
average over the whole sky, which in practice means a signal on scales
larger than about two degrees - the signal on these scales is expected to
be the same in all directions on the sky.

 However, by far the best opportunity to measure the cosmological signal
(again thanks to the spectral smoothness of the foregrounds) is offered
by its frequency structure. The large-scale density fluctuations in a
single beam are uncorrelated on frequency scales above about
$0.2-0.3\dim{MHz}$, so that the excess brightness temperature varies
by about $2-3\dim{mK}$ and the local spectral index of the radio
signal varies by about 0.4\% on scales of the order of $1\dim{MHz}$.
These sharp variations should not be difficult to separate from the smoothly
varying foreground signal.

Thus, the frequency fluctuation signal stands out in comparison with the
other two possibilities (mean signal and angular fluctuations). While
the observability of the latter two depends critically on confusion with
foregrounds, the former may be largely just a matter of sensitivity.
Frequency calibration, contamination by spectral lines or leakage of
angular fluctuations, and RFI may all be manageable problems. Using our
simulations we have shown that it may then be possible to detect the
$\HI$ signal from the pre-reionization era even at moderate angular
resolution ($\sim10-20$ arcmin), corresponding to filled apertures of a few
hundred meters diameter.

\acknowledgements
This work was supported in part by NSF grant AST-0134373 and by
National Computational Science Alliance under grant AST-020018N and
utilized SGI Origin 2000 array and IBM P690 array at the National Center
for Supercomputing Applications.

\end{document}